\documentclass[12pt,letterpaper]{article}

\usepackage{lineno,amsmath}
\usepackage{pifont,geometry,graphicx,lineno,amssymb,xspace}
\usepackage{framed}
\usepackage{placeins}
\usepackage[usenames,dvipsnames]{xcolor}
\usepackage[colorlinks=true, linkcolor=darkgray, citecolor=darkgray]{hyperref}
\usepackage{amsfonts}
\usepackage{natbib}
\usepackage{listings}
\usepackage{color}
\usepackage{widetext,array}
\usepackage{setspace}
\usepackage{mdframed}
\usepackage{authblk}

\newcommand{\ave}[1]{\left\langle#1 \right\rangle}

\newcommand{\elabel}[1]{\label{eq:#1}}
\newcommand{\eref}[1]{equation~(\ref{eq:#1})}
\newcommand{\Eref}[1]{Equation~(\ref{eq:#1})}

\newcommand{\flabel}[1]{\label{fig:#1}}
\newcommand{\fref}[1]{Figure~\ref{fig:#1}}
\newcommand{\Fref}[1]{Figure~\ref{fig:#1}}

\newcommand{\U}{\gls{U}}
\newcommand{\Ui}{\gls{U_ins}}
\newcommand{\Uo}{\gls{U_own}}
\newcommand{\W}{\gls{W}}
\newcommand{\Wo}{\gls{W_own}}
\newcommand{\Wi}{\gls{W_in}}
\newcommand{\C}{\gls{C}}
\newcommand{\F}{\gls{F}}
\newcommand{\G}{\gls{G}}
\newcommand{\p}{\gls{p}}

\newcommand{\ie}{{\it i.e.}~}
\newcommand{\eg}{{\it e.g.}~}

\usepackage{glossaries}
\newglossary{symbols}{sym}{sbl}{List of Symbols}
\makenoidxglossaries
\newglossaryentry{rex}
{
 type={symbols}, 
 name={\ensuremath{\ave{r}}},
 description={Rate of change of the expectation value. In discrete time, considered here, this is also the expectation value of the rate of change because both the differencing in computing the rate of change and the summing in computing the expectation value are linear operations that commute. In continuous time we have to be more careful because the rate of change $\frac{\Delta W}{\Delta t}$ may not exist in the limit $\Delta t \to 0$},
 sort=rex
}

\newglossaryentry{Delta_U}
{
 type={symbols}, 
 name={\ensuremath{\Delta U}},
 description={Change in utility over one round trip},
 sort=Delta_U
}

\newglossaryentry{Delta_W}
{
 type={symbols}, 
 name={\ensuremath{\Delta W}},
 description={Change in wealth over one round trip},
 sort=Delta_W
}

\newglossaryentry{delta_r_own}
{
 type={symbols}, 
 name={\ensuremath{\delta \ave{r}_{\text{own}}}},
 description={Change in the rate of change of the expectation value of wealth, experienced by the shipowner when the insurance contract is signed},
 sort=delta_r_own
}

\newglossaryentry{Delta_r_own}
{
 type={symbols}, 
 name={\ensuremath{\Delta \ave{r}_{\text{own}}}},
 description={Change in the rate of change of the expectation value of wealth, experienced by the shipowner when the insurance contract is signed},
 sort=delta_r_own
}

\newglossaryentry{delta_r_ins}
{
 type={symbols}, 
 name={\ensuremath{\delta \ave{r}_{\text{ins}}}},
 description={Change in the rate of change of the expectation value of wealth, experienced by the insurer when the insurance contract is signed},
 sort=delta_r_ins
}

\newglossaryentry{Delta_r_ins}
{
 type={symbols}, 
 name={\ensuremath{\Delta \ave{r}_{\text{ins}}}},
 description={Change in the rate of change of the expectation value of wealth, experienced by the insurer when the insurance contract is signed},
 sort=delta_r_ins
}

\newglossaryentry{r_own_un}
{
  type={symbols}, 
  name={\ensuremath{\ave{r}_{\text{own}}^{\text{un}}}},
  description={Rate of change of the expectation value of the uninsured shipowner's wealth},
  sort=rownun
}

\newglossaryentry{r_own_in}
{
  type={symbols}, 
  name={\ensuremath{\ave{r}_{\text{own}}^{\text{in}}}},
  description={Rate of change of the expectation value of the insured shipowner's wealth},
  sort=rownin
}

\newglossaryentry{ru}
{
  type={symbols}, 
  name={\ensuremath{\ave{r_u}}},
  description={Rate of change of the expectation value of utility},
  sort=rinun
}

\newglossaryentry{r_in_un}
{
  type={symbols}, 
  name={\ensuremath{\ave{r}_{\text{ins}}^{\text{un}}}},
  description={Rate of change of the expectation value of the insurer's wealth without insurance contract},
  sort=rinun
}

\newglossaryentry{r_in_in}
{
  type={symbols}, 
  name={\ensuremath{\ave{r}_{\text{ins}}^{\text{in}}}},
  description={Rate of change of the expectation value of the insurer's wealth with insurance contract},
  sort=rinin
}

\newglossaryentry{Delta_ru_own}
{
 type={symbols}, 
 name={\ensuremath{\Delta \ave{r_u}_{\text{own}}}},
 description={Change in the rate of change of the expectation value of utility, experienced by the shipowner when the insurance contract is signed},
 sort=delta_ru_own
}

\newglossaryentry{delta_ru_own}
{
 type={symbols}, 
 name={\ensuremath{\delta \ave{r_u}_{\text{own}}}},
 description={Change in the rate of change of the expectation value of utility, experienced by the shipowner when the insurance contract is signed},
 sort=delta_ru_own
}

\newglossaryentry{delta_ru_ins}
{
 type={symbols}, 
 name={\ensuremath{\delta \ave{r_u}_{\text{ins}}}},
 description={Change in the rate of change of the expectation value of utility, experienced by the insurer when the insurance contract is signed},
 sort=delta_ru_ins
}

\newglossaryentry{Delta_ru_ins}
{
 type={symbols}, 
 name={\ensuremath{\Delta \ave{r_u}_{\text{ins}}}},
 description={Change in the rate of change of the expectation value of utility, experienced by the insurer when the insurance contract is signed},
 sort=delta_ru_ins
}

\newglossaryentry{ru_own_un}
{
  type={symbols}, 
  name={\ensuremath{\ave{r_u}_{\text{own}}^{\text{un}}}},
  description={Rate of change of the expectation value of the uninsured shipowner's utility},
  sort=ruownun
}

\newglossaryentry{ru_own_in}
{
  type={symbols}, 
  name={\ensuremath{\ave{r_u}_{\text{own}}^{\text{in}}}},
  description={Rate of change of the expectation value of the insured shipowner's utility},
  sort=ruownin
}

\newglossaryentry{ru_in_un}
{
  type={symbols}, 
  name={\ensuremath{\ave{r_u}_{\text{ins}}^{\text{un}}}},
  description={Rate of change of the expectation value of the insurer's utility without insurance contract},
  sort=ruinun
}

\newglossaryentry{ru_in_in}
{
  type={symbols}, 
  name={\ensuremath{\ave{r_u}_{\text{ins}}^{\text{in}}}},
  description={Rate of change of the expectation value of the insurer's utility with insurance contract},
  sort=ruinin
}

\newglossaryentry{g_time}
{
  type={symbols}, 
  name={\ensuremath{\bar{g}}},
  description={Time-average growth rate},
  sort=g_bar
}

\newglossaryentry{delta_g_own}
{
 type={symbols}, 
 name={\ensuremath{\delta \bar{g}_{\text{own}}}},
 description={Change in the time average (and expectation value) of the dynamic-specific growth rate of wealth (here exponential), experienced by the shipowner when the insurance contract is signed},
 sort=delta_g_own
}

\newglossaryentry{delta_g_ins}
{
 type={symbols}, 
 name={\ensuremath{\delta \bar{g}_{\text{ins}}}},
 description={Change in the time average (and expectation value) of the dynamic-specific growth rate of wealth (here exponential), experienced by the insurer when the insurance contract is signed},
 sort=delta_g_ins
}

\newglossaryentry{Delta_g_own}
{
 type={symbols}, 
 name={\ensuremath{\Delta \bar{g}_{\text{own}}}},
 description={Change in the time average (and expectation value) of the dynamic-specific growth rate of wealth (here exponential), experienced by the shipowner when the insurance contract is signed},
 sort=delta_g_own
}

\newglossaryentry{Delta_g_ins}
{
 type={symbols}, 
 name={\ensuremath{\Delta \bar{g}_{\text{ins}}}},
 description={Change in the time average (and expectation value) of the dynamic-specific growth rate of wealth (here exponential), experienced by the insurer when the insurance contract is signed},
 sort=delta_g_ins
}

\newglossaryentry{g_own_un}
{
  type={symbols}, 
  name={\ensuremath{\bar{g}_{\text{own}}^{\text{un}}}},
  description={Time-average growth rate of the uninsured shipowner's wealth},
  sort=gownun
}

\newglossaryentry{g_own_in}
{
  type={symbols}, 
  name={\ensuremath{\bar{g}_{\text{own}}^{\text{in}}}},
  description={Time-average growth rate of the insured shipowner's wealth},
  sort=gownin
}

\newglossaryentry{g_in_un}
{
  type={symbols}, 
  name={\ensuremath{\bar{g}_{\text{ins}}^{\text{un}}}},
  description={Time-average growth rate of the insurer's wealth without insurance contract},
  sort=ginun
}

\newglossaryentry{g_in_in}
{
  type={symbols}, 
  name={\ensuremath{\bar{g}_{\text{ins}}^{\text{in}}}},
  description={Time-average growth rate of the insurer's wealth with insurance contract},
  sort=ginin
}

\newglossaryentry{U}
{
  type={symbols}, 
  name={\ensuremath{U}},
  description={Utility function},
  sort=U
}

\newglossaryentry{U_own}
{
  type={symbols}, 
  name={\ensuremath{U_{\text{own}}}},
  description={Shipshipowner's utility function},
  sort=U
}

\newglossaryentry{U_ins}
{
  type={symbols}, 
  name={\ensuremath{U_{\text{ins}}}},
  description={Insurer's utility function},
  sort=U
}

\newglossaryentry{W}
{
  type={symbols}, 
  name={\ensuremath{W}},
  description={Wealth},
  sort=W
}

\newglossaryentry{W_own}
{
  type={symbols}, 
  name={\ensuremath{W_{\text{own}}}},
  description={Wealth of the insurance buyer},
  sort=wown
}

\newglossaryentry{W_in}
{
  type={symbols}, 
  name={\ensuremath{W_{\text{ins}}}},
  description={Wealth of the insurer},
  sort=wins
}

\newglossaryentry{C}
{
type={symbols},
  name={\ensuremath{C}},
  description={Replacement cost of the ship},
  sort=C
}

\newglossaryentry{F}
{
  type={symbols}, 
  name={\ensuremath{F}},
  description={Insurance fee},
  sort=F
}

\newglossaryentry{G}
{
  type={symbols}, 
  name={\ensuremath{G}},
  description={Gain from one round trip of the ship},
  sort=G
}

\newglossaryentry{p}
{
  type={symbols}, 
  name={\ensuremath{p}},
  description={Probability of losing the ship on one round trip},
  sort=p
}

\newglossaryentry{t}
{
  type={symbols}, 
  name={\ensuremath{t}},
  description={Time when the goods leave Amsterdam},
  sort=t
}

\newglossaryentry{L}
{
  type={symbols}, 
  name={\ensuremath{L}},
  description={Insured loss},
  sort=L
}

\newglossaryentry{Delta_t}
{
  type={symbols}, 
  name={\ensuremath{\Delta t}},
  description={Duration of one round trip},
  sort=Deltat
}

\begin{document}

\title{Insurance makes wealth grow faster}
\author[1,2]{Ole Peters\thanks{o.peters@lml.org.uk}}
\author[1]{Alexander Adamou\thanks{a.adamou@lml.org.uk (corresponding author)}}
\affil[1]{London Mathematical Laboratory, 14 Buckingham Street, London WC2N 6DF, UK}
\affil[2]{Santa Fe Institute, 1399 Hyde Park Road, 87501 Santa Fe, NM, USA}
\date{\today}
\maketitle


\begin{abstract}

Voluntary insurance contracts constitute a puzzle because they increase the expectation value of one party's wealth, whereas both parties must sign for such contracts to exist. Classically, the puzzle is resolved by introducing non-linear utility functions, which encode asymmetric risk preferences; or by assuming the parties have asymmetric information. Here we show the puzzle goes away if contracts are evaluated by their effect on the time-average growth rate of wealth. Our solution assumes only knowledge of wealth dynamics. Time averages and expectation values differ because wealth changes are non-ergodic. Our reasoning is generalisable: business happens when both parties grow faster.
\end{abstract}

\newpage

\section{Introduction}

\subsection{Conceptual background}
This study was motivated by an observation by Kenneth Arrow, articulated at the
30-year anniversary symposium of the Santa Fe Institute in 2014.
Arrow listed as one significant weakness of general competitive equilibrium theory its inability to
provide a satisfactory answer to the question: why do insurance contracts exist? Since financial derivatives -- 
a form of insurance -- constitute the largest market on earth, the existing answers, he claimed, 
are not good enough. Even in the modern insurance literature, explaining the existence of insurance contracts has been identified as central to {\it ``explain[ing] the `raison d'\^etre' of the insurance industry''} \citep[p.~241]{vanCayseele1992}.

The conventional economic explanations for why insurance exists are asymmetric information, risk aversion, and irrationality. Here we provide an alternative explanation whose basis is dynamics: activities resembling insurance are likely to emerge whenever multiple entities are faced with managing resources
in an environment of noisy non-additive growth.

In the 17$^\text{th}$ century people concerned with economic problems developed 
probabilistic mathematical reasoning \citep{Devlin2008}. Part of this 
innovation was the invention of the expectation value.
It was explicitly assumed that an individual will make decisions to optimise the rate of 
change of the expectation value of his wealth. We call this model of human behaviour 
the ``expected-wealth paradigm.'' Implicit in this paradigm is the ergodic hypothesis
that the expectation value reflects what happens over a long time.\footnote{A
rare example of this belief being stated explicitly in the context of decision theory is in \citep[p.98]{ChernoffMoses1959}: 
{\it ``If a gamble is favorable from the point of view of the expectation value and you have the 
choice of repeating it many times, then it is wise to do so. For eventually, your amount of 
money [is] bound to increase.''} The ergodic hypothesis is more frequently explicit in the study of economic inequality, see \citep{BermanETAL2017} and references therein.} 
We now know that this equivalence holds only under special conditions, the study of which is
known as ergodic theory. Growth processes generally violate these conditions. A comprehensive discussion of the ergodic hypothesis in decision theory can be found in \citep{PetersGell-Mann2016}.

It is helpful to delineate the physical meaning of the expectation value. It 
is the average over an infinitely large ensemble of identically prepared systems, real or imagined, that evolve randomly into the future. 
This characterisation motivates the question: why would an individual optimise such a quantity? 
\textit{A priori} the many systems embedded in the expectation value are not accessible to the individual, so there is no good reason 
for him to maximise it. This constitutes a conceptual challenge of the axioms of the 
expected-wealth paradigm. 

Historically, the expected-wealth paradigm was first challenged
by N.~Bernoulli in his famous 1713 letter to Montmort \citep{Montmort1713}, whose contents 
became known as the ``St. Petersburg 
paradox.'' The conclusion was inescapable: human behaviour does not optimise the 
expectation value of wealth. This is an empirical challenge to the paradigm. 

The challenge was answered by the development of the expected utility hypothesis by Cramer and 
D.~Bernoulli \citep{Bernoulli1738}. 
It aims to capture the difference between observed behaviour and that predicted by expected 
wealth maximisation by converting money non-linearly into ``utility.'' The conversion is specified so 
that people maximise the expectation value of utility. Note that the key axiom -- 
that an expectation value is being maximised -- remained in place. A utility function
was sought that would describe human behaviour within the existing conceptual space, 
and the resulting formalism is first and foremost descriptive and phenomenological.
The conceptual maturity of expected-utility theory is limited by the concepts available
at the time of its development.\footnote{We consider the 1944 axiomatisation of expected 
utility theory \citep{vonNeumannMorgenstern1944} a mathematical refinement rather than a conceptual renewal of the original ideas.} 
Crucially, these concepts do not include the question central in ergodic theory, whether expectation values reflect what happens in reality over a long time.

We question directly 
the meaningfulness of expectation values by dropping the ergodic hypothesis. 
Since we identify what happens over time
as the relevant question, for the remainder of this paper we call our approach
the ``time paradigm.''
 
Some 200 years after the invention of quantitative probabilistic reasoning, the 
``ergodicity question'' arose: do
expectation values reflect what happens over time? Are the ups and downs in a fluctuating 
quantity equivalent in the long run to the quantity sitting at its expectation value? 
The well-known answer is no. In general, if we're interested in the long-time effect of 
fluctuations we must explicitly include time
in our models and compute time-averages of the quantities of interest. Only in 
special cases, in which these quantities are ergodic, can we rely on expectation values being indicative of 
long-time behaviour.

This is significant in many economic problems. Economics tends to deal with 
growth processes, for example in personal wealth or national income. Growing quantities cannot be ergodic but it may be possible to construct observables, such as growth rates, whose time averages equal their expectation 
values \citep{PetersGell-Mann2016}. 
This possibility is not guaranteed: the relevant ergodic properties must be 
demonstrated, not assumed, for each observable we wish to analyse
with methods reliant on ergodicity.

Intuition resting on the assumption that things behave over time just as they behave in a
statistically stationary ensemble is misleading here.
As we shall see in the present context, this flawed intuition makes insurance contracts
appear as zero-sum games that benefit only one party. By eliminating the ergodic assumption, 
insurance contracts -- and by extension other business deals -- appear as win-win situations
that benefit both parties.

\subsection{The insurance puzzle and two treatments}

We revisit the question of why insurance contracts exist, unresolved in
the expected-wealth paradigm. 
The issues at hand are ostensibly about insurance, interesting from the perspectives of 
both actuarial science and decision theory. Through decision theory their relevance 
extends to much of formal economics. For instance, the trivial extension from insurance 
contracts to financial derivatives demonstrates the significance of the time paradigm for 
modern financial markets. 

\subsubsection{The problem with insurance in the expected-wealth paradigm}
\label{The_insurance}
Let us be clear about what the expected-wealth paradigm is.

\begin{center}
{\it The expected-wealth 
paradigm is a model of human behaviour. It posits that 
humans act to maximise the rate of change of the expectation value of their wealth.}
\end{center}

It predicts the absence of insurance contracts as follows:
\begin{enumerate}
\item
To be profitable, an insurer has to charge an insurance premium greater than the expectation value of any claims that may be made against 
it, called the ``net premium,'' \citep[p.~1]{KaasETAL2008}.
\item
The insurance buyer therefore has to be willing to pay more than the
net premium so that an insurance contract may be successfully signed.
\item
Under the expected-wealth paradigm it is irrational for the buyer to pay more 
than the net premium, and therefore insurance contracts should 
not exist.
\end{enumerate}
An insurance contract can only ever be beneficial to one party. It has the anti-symmetric property that
the expectation value of the gain of one party is the expectation value of 
the loss of the other party. From this perspective, it is a zero-sum game.

\subsubsection{Treatment 1 -- the utility paradigm}
\citep{Bernoulli1738} identified the rationality model in step 3 above as 
problematic and devised the utility paradigm.  
He observed that money 
may not translate linearly into usefulness and assigned to any individual 
an idiosyncratic utility function $\U(\W)$ that maps wealth $\W$ into 
usefulness $\U$, which he posited as the true quantity whose expected change 
is maximised in a decision. 

\begin{center}
{\it The utility paradigm is another model of human behaviour. 
It posits that humans act to maximise the rate of change of the expectation value of their utility.}
\end{center}

Since the utility function of the insurance buyer may be 
non-linear, it is possible that the buyer experiences a positive change in the 
expectation value of his utility from signing an insurance contract, even if the 
expectation value of his monetary wealth decreases. Therefore, the 
utility paradigm does not rule out the existence of insurance contracts. It does 
not share the anti-symmetric property of the expected-wealth paradigm. 
The expectation value of the gain of utility of one party need not be the 
expectation value of the loss of utility of the other party.

In fact, since different people have different wealths or different
utility functions or both, the change in the expectation value of utility experienced
by one party is almost unrestricted by the change experienced by 
the other party. This is problematic and led \citet[p.~918]{Kelly1956} 
to comment that the utility paradigm is {\it ``too general to shed any light on the 
specific problems''}. 
The paradigm affords too much freedom by appealing to 
individual differences 
and by allowing too broad a class of utility functions. As 
in any paradigm, generality comes at the expense of predictive power. In
the case of the utility paradigm, the predictive power is often limited to the 
point of practical uselessness.  Utility theory dominates formal economics but plays
no major role in practical actuarial work, where limiting the likelihood of the insurer's insolvency is the paradigmatic idea.\footnote{This approach, known as ruin theory, originated outside economics at the start of the $20^\text{th}$ century \citep{Lundberg1903} and remained there. Indeed, \citet[p.~134]{Arrow1971} remarked that {\it ``Insurance is an item of considerable importance in the economies of advanced nations, yet it is fair to say that economic theorists have had little to say about it, and insurance theory has developed with virtually no reference to the basic economic concepts of utility and productivity.''}}

\subsubsection{Treatment 2 -- the time paradigm}
In agreement with Bernoulli we take issue with step 3 of Section \ref{The_insurance}. 
It does indeed imply an inappropriate model of rationality.  But the 
correction we propose is entirely different from Bernoulli's in its 
epistemology. 
\begin{center}
{\it
The time paradigm is another model of human behaviour. It posits that humans act 
to maximise the long-time rate of change of their wealth.
}
\end{center}

The time-average 
growth rate of wealth under reasonable dynamics (the standard being multiplicative 
dynamics) behaves very differently from the rate of change of the expectation value of wealth.
This is because the appropriate growth rate for a given dynamic must take into account the
non-linearity of the process. 
Typically there exists a 
range of prices where time-average growth rates increase for both parties when 
an insurance contract is signed. From this perspective, reasonably priced insurance contracts are win-win deals.

Without appealing to utility functions, the time paradigm predicts the 
existence of insurance contracts with deep conceptual consequences. 

\section{Theory and calculation}

\subsection{A shipping example}
We use the most basic model of an insurance contract, considered \eg by 
\citep{RothschildStiglitz1976}. The model is chosen because it is simple and 
has all the necessary elements to demonstrate the failure of the expected-wealth 
paradigm and the different fixes under the utility paradigm and under 
the time paradigm. \citep{Bernoulli1738} applied the model in the context of 
shipping, and we will do the same for illustration.\footnote{The main focus 
of \citep{Bernoulli1738}, the St.\ Petersburg paradox, was
resolved using the time paradigm in \citep{Peters2011a}.}
To make the model more concrete and demonstrate that 
it works in a practically interesting regime, we use example values for its
parameters.

A St.\ Petersburg merchant sends a ship to Amsterdam. Upon the safe arrival of the ship the 
merchant will make a gain of $\G=\$4,000$. However, 
with probability $\gls{p}=0.05$ the ship, whose replacement cost is $\C=\$30,000$, will be lost. 
The journey is known to take $\gls{Delta_t}=1$~month. The shipowner's wealth when 
the ship sets sail at time $t$ is $\gls{W_own}=\$100,000$. An insurer proposes a 
contract stipulating that, should the ship be lost, the shipowner shall receive 
the replacement cost of the ship and his lost profit, $\gls{L}=\G+\C$, 
for an insurance fee of $\gls{F}=\$1,800$. If the two parties sign the contract the shipowner 
will receive $\gls{G}-\F$ with certainty after one month, whereas the insurer will 
receive $\F$ with probability $1-\gls{p}$ and lose $\gls{L}-\F$ with probability $\gls{p}$. 
The contract is completely specified by the insured loss $\gls{L}$ and the 
fee $\gls{F}$ (and the payout condition, of course). We assume that both 
parties have perfect information, \ie both know the true probability of the 
ship being lost, and that there is no possibility of fraud or default. Should the shipowner sign 
the insurance contract, and did the insurer act in his own interest by proposing it?

\subsection{The expected-wealth paradigm and its problems}
\label{The_expectation}

The expected-wealth paradigm, in the language of economics, assumes 
that humans are risk neutral, \ie they have no preference between gambles 
whose expected changes in wealth over a given time interval are identical. 
This assumption has been known to 
be flawed at least since 1713 \citep[p.~402]{Montmort1713}. We discuss it here to 
better understand the origin of the utility paradigm and enable a conceptually 
different approach. Under the expected-wealth paradigm humans act 
to maximise the rate of change of the expectation value of their wealth:\footnote{We denote by $\ave{x}$ the expectation value of a random variable, $x$,  and define it as the average over infinitely many independent realistions: $\ave{x} \equiv \lim_{N\rightarrow\infty} \left\{ \frac{1}{N} \sum_{i=1}^N x_i\right\}$.}
\begin{equation}
\gls{rex}=\frac{\ave{\gls{Delta_W}}}{\gls{Delta_t}}=\frac{\ave{\W(\gls{t}+\gls{Delta_t})}-\W(\gls{t})}{\gls{Delta_t}}.
\elabel{expectation_paradigm}
\end{equation}
The attractiveness of an insurance contract is judged by
computing the change in \gls{rex} that results from signing the contract.\\

\noindent\underline{The shipowner's perspective}\\

{\it Without}
insurance, the rate of change of the expectation value of the shipowner's wealth is 
\begin{equation}
\gls{r_own_un}=\frac{(1-\gls{p})\gls{G}-\p\C}{\gls{Delta_t}}=\frac{\gls{G}-\p\gls{L}}{\gls{Delta_t}},
\end{equation} 
or $\$2,300$ per month using the example parameters.

{\it With} insurance the certain rate of change of the shipowner's wealth is
\begin{equation}
\gls{r_own_in}=\frac{\G-\gls{F}}{\gls{Delta_t}}. 
\end{equation} 
In the example this is $\$2,200$ per month. 

The change in the rate of change of the
expectation value of the shipowner's wealth resulting from entering into the insurance contract is
\begin{equation}
\gls{delta_r_own}=\gls{r_own_in}-\gls{r_own_un}=\frac{\gls{p} \gls{L}-\gls{F}}{\gls{Delta_t}},
\elabel{delta_r_own}
 \end{equation} 
here $-\$100$ per month. This change is marked by $\delta$ rather than $\Delta$ to clarify that it is not the change over $\gls{Delta_t}$ of a time-varying quantity. 

Buying insurance in this example reduces the rate of change of the
expectation value of the shipowner's wealth. Hence, the shipowner
should not sign the contract according to the expected-wealth paradigm.\\

\noindent\underline{The insurer's perspective}\\

{\it Without} insurance, the
certain rate of change of the insurer's wealth is
\begin{equation} 
\gls{r_in_un}=0,
\end{equation} 
since the insurer does no business.

{\it With} insurance, the rate of change of the expectation value of the insurer's wealth is 
\begin{equation} 
\gls{r_in_in}=\frac{\gls{F}-\gls{p}\gls{L}}{\gls{Delta_t}}, 
\elabel{r_in_in}
\end{equation} 
or $+ \$ 100$ per month in the example.

The change in the rate of change of the expectation value of the
insurer's wealth resulting from entering into the contract is 
\begin{equation}
\gls{delta_r_ins}=\gls{r_in_in}-\gls{r_in_un}=\frac{\gls{F}-\gls{p}\gls{L}}{\gls{Delta_t}},
\elabel{delta_r_ins}
\end{equation}
here $+\$100$ per month. Hence the insurer should
sign the contract  according to the expected-wealth paradigm. But since the shipowner should 
not sign, no deal will be made.

Crucially, irrespective of the parameter values, what the insurer gains is always 
precisely what the shipowner loses. Comparing \eref{delta_r_own} and \eref{delta_r_ins}, we see that
\begin{equation}
\gls{delta_r_ins}=-\gls{delta_r_own}.
\end{equation}
This anti-symmetry makes insurance a fundamentally unsavoury business -- a zero-sum game 
where one party wins at the expense of the other. 
The existence of such contracts in the real world requires asymmetries 
between the contracting parties: buyer and seller may have different 
access to information; they may make different subjective assessments 
of the risk being insured (encoded in their estimates of model parameters 
such as $\gls{p}$ and $\gls{L}$); or they may deceive, coerce, or gull the other 
party into the necessary sub-optimal decision. We consider this

\mbox{}
\newpage
\begin{mdframed}[linewidth=2pt]
\underline{The problem with the expected-wealth paradigm:}\\
\\
\textbf{No price exists that makes an insurance contract beneficial to both parties, and yet insurance contracts exist.}
\end{mdframed}
\mbox{}

We summarise the two perspectives in Table~\ref{expectation_table} and
illustrate the effect in \fref{expectation_fee}.

\begin{table}
\label{expectation_table}
\noindent\makebox[\textwidth]{
\begin{tabular}{ m{2cm} m{4cm} m{5cm} }
\gls{rex} & Shipowner & Insurer \\
\hline
insured & $\gls{r_own_in}=\frac{\gls{G}-\gls{F}}{\gls{Delta_t}}$ &$\gls{r_in_in}=\frac{\gls{F}-\gls{p}\gls{L}}{\gls{Delta_t}}$\\
uninsured &$\gls{r_own_un}=\frac{(1-\gls{p})\gls{G}-\p\C}{\gls{Delta_t}}$ &$\gls{r_in_un}=0$\\
\hline
difference &$\gls{delta_r_own}=\frac{\gls{p}\gls{L}-\gls{F}}{\gls{Delta_t}}$ &$\gls{delta_r_ins}=-\gls{delta_r_own}=\frac{\gls{F}-\gls{p}\gls{L}}{\gls{Delta_t}}$\vspace{.1cm}
\\
\hline
\hline
\end{tabular}
}
\caption{Changes in the rates of change of the expectation value of wealth for shipowner and insurer. None of the expressions contains the wealth of either party.}
\end{table}

\begin{figure}
\setlength{\unitlength}{\textwidth}
\begin{picture}(1,0.72)
  \put(0,0){\includegraphics[width=1.\textwidth]{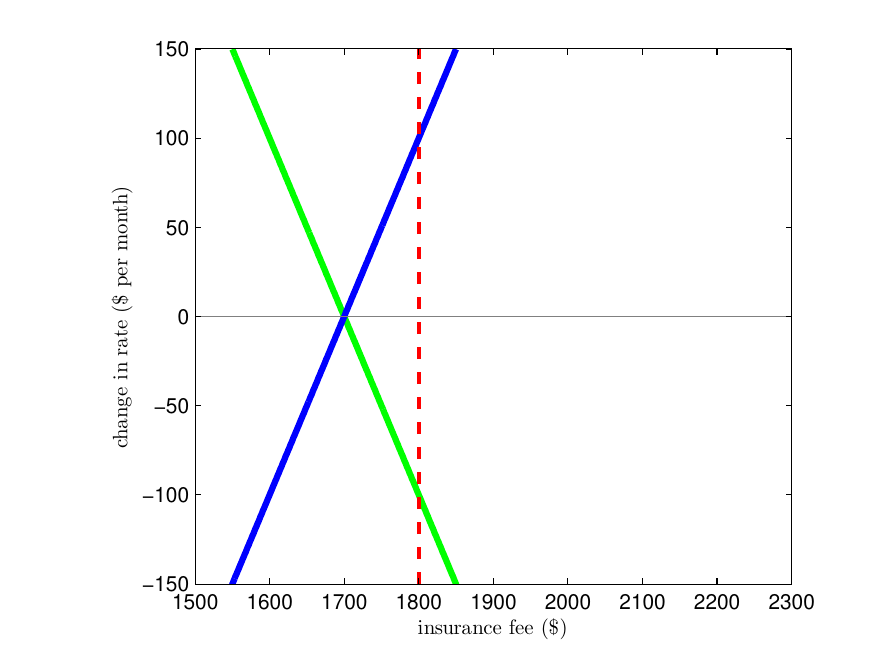}}
  \put(0.55,0.42){\includegraphics[width=0.35\textwidth]{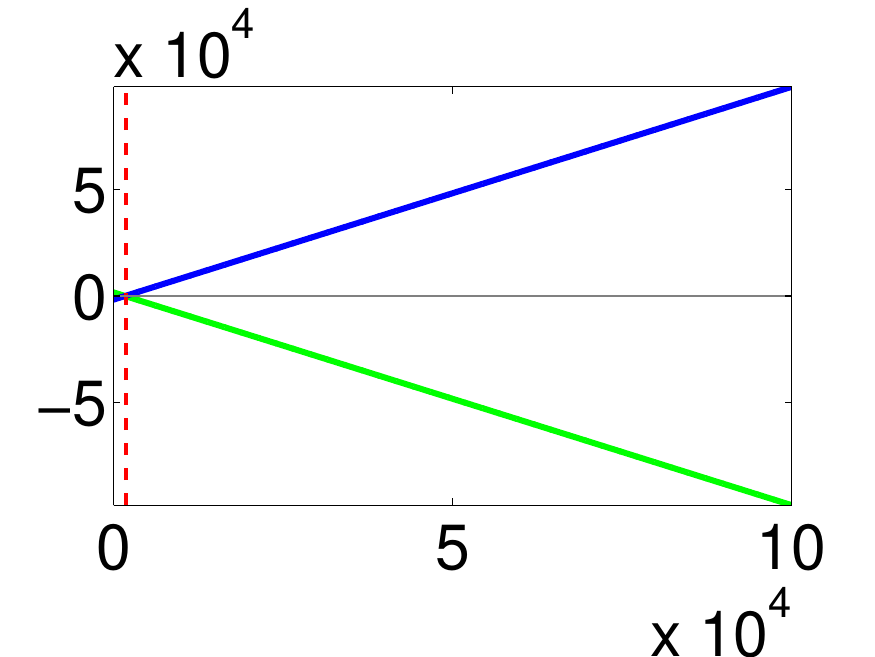}}
\end{picture}
\caption{Change in the rate of change of the expectation value of wealth resulting from signing an insurance
  contract versus insurance fee.  \underline{Solid line:} shipowner. \underline{Dashed line:} insurer. 
  The straight lines of opposite slopes 
  cross at zero, meaning that there is no price where both lines lie above 
  the zero line -- the winning party only gains at the expense of the losing party. The vertical 
  dotted line indicates the proposed insurance fee, which appears attractive to the insurer 
  and unattractive to the shipowner. \underline{Inset:} horizontal range extended to \$100,000.}
\flabel{expectation_fee}
\end{figure}


\subsection{Solution in the utility paradigm}
\label{Solution_in_the_utility}
The model of human decision making of the expected-wealth paradigm
predicts that insurance contracts do not exist. This is in disagreement 
with observations. D.~Bernoulli concluded 
that humans use a different criterion when deciding whether to sign
insurance contracts \citep{Bernoulli1738}. He introduced a non-linear mapping of money, 
declaring: if humans don't act to maximise the expectation value of money, 
then let's say they act to maximise the expectation value of some sub-linearly 
increasing function of money. He called this function the ``utility function'' and
left it largely unspecified, using the logarithm in 
computations
and pointing out that it produces essentially the same results as a square root 
function.\footnote{\citep{Bernoulli1738} is credited with computing changes 
in the expectation values of utility functions although a careful reading of his
work shows that he computed something slightly different \citep{Peters2011a}. The 
difference was small in his context, and later researchers assumed he had meant 
to compute changes in expectation values \citep[p.~439--442]{Laplace1814}.}  
In effect, the idiosyncrasies of human beings and their circumstances were invoked. A dollar to
a rich man has a value different from a dollar to a poor man; some
people like gambling, others are more prudent.

This approach does not so
much resolve the puzzle as provide a description in mathematical
terms that is consistent with a description in everyday language: the 
expected-wealth paradigm does not work. The introduction of a non-linear 
utility function breaks the equality of the locations of 
zero-crossings of the value of the insurance contract for both parties. It is this 
equality that makes insurance impossible in the 
expected-wealth paradigm. Any perturbation that breaks 
it and creates a win-win range of fees suffices to solve the 
mathematical problem. Introducing non-linear utility functions, 
perhaps different for shipowner and insurer, constitutes 
such a perturbation. An example is shown in \fref{utility_fee}, where utility is 
given by the square-root of wealth, $\U(\W)=\sqrt{\W}$ (the first ever utility
function to be suggested, by Cramer in a 1728 letter cited by \citep{Bernoulli1738}).
A more satisfying resolution of the puzzle would explain the rationale behind
such a perturbation. Why a square-root? Why a logarithm? Why re-value
money at all?

\begin{figure}
\setlength{\unitlength}{\textwidth}
\begin{picture}(1,0.72)
  \put(0,0){\includegraphics[width=1.\textwidth]{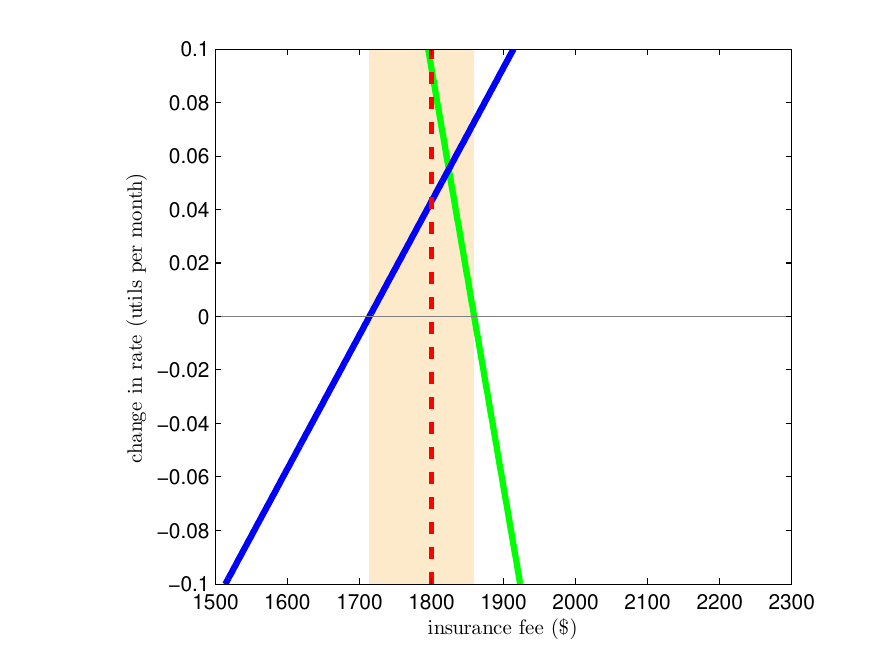}}
  \put(0.55,0.42){\includegraphics[width=0.35\textwidth]{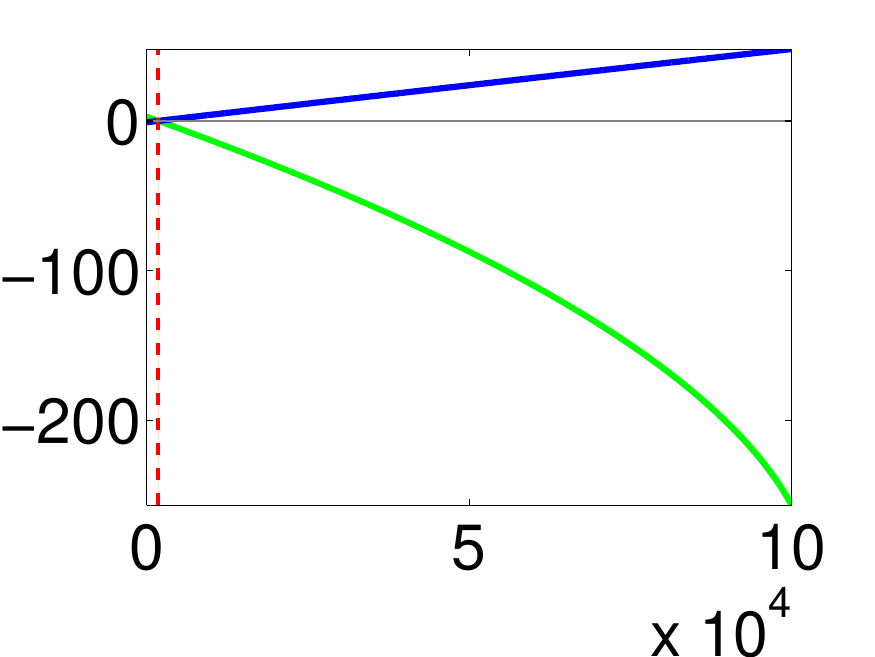}}
\end{picture}
\caption{Change in the rate of change of the expectation value of square-root utility
  resulting from signing an insurance contract versus insurance fee.  \underline{Solid line:} shipowner. 
  \underline{Dashed line:} insurer. The vertical dotted line indicates the proposed insurance fee. 
  \underline{Shaded region}: Introducing the non-linear utility function is a sufficient perturbation to create a 
  regime where both parties gain. Both the solid and dashed lines are above the zero line here. 
  However, the non-linear function is ill-constrained by the formalism, which
  limits predictive power. \underline{Inset:} horizontal axis extended to \$100,000. The non-linearity
  is clearly visible at these scales.} \flabel{utility_fee}
\end{figure}

In this paradigm we consider the rate of change of the expectation value of utility, 
\begin{equation}
\gls{ru}=\frac{\ave{\gls{Delta_U}}}{\gls{Delta_t}}=\frac{\ave{\U(\gls{t}+\gls{Delta_t})}-\U(\gls{t})}{\gls{Delta_t}}.
\elabel{ru}
\end{equation} 

Again using the shipping example, the equations corresponding to those in
Section~\ref{The_expectation} are as follows.\\

\noindent\underline{The shipowner's perspective}\\

{\it Without} insurance, the rate of change of the expectation value of the shipowner's utility is
\begin{equation}
\gls{ru_own_un}=\frac{(1-\p) \Uo(\Wo+\gls{G})+\p \Uo(\Wo-\C)- \U(\Wo)}{\gls{Delta_t}}.
\end{equation} 
This quantity is measured in ``utils'' per unit time, with the dimension util 
dependent on the utility function.\footnote{The fact that the util is a dimension imposes
restrictions on the utility function that are not generally recognised. 
\citet[p.~17 ff.]{Barenblatt2003} shows that any dimension function of any physically 
meaningful quantity, such as money, must be a power-law monomial, which 
restricts utility functions themselves to power-law monomials. This is a failure of the utility
paradigm as it is often presented. The interpretation 
of $\ln(\W)$ alone as a physically meaningful quantity is untenable -- it should only ever
occur as $\ln(\W_1)-\ln(\W_2)=\ln(\W_1/\W_2)$, where utils cancel out and the 
logarithm is taken of  dimensionless quantities only. We note that 
\citet{vonNeumannMorgenstern1944} pointed out that only differences 
between utilities may be considered meaningful, which avoids the problem.} With the example 
parameter values and square-root utility this is 3.37 utils per month.

{\it With} insurance the (certain) rate of change of utility is
\begin{equation}
\gls{ru_own_in}=\frac{\Uo(\Wo+\gls{G}-\F)-\Uo(\Wo)}{\gls{Delta_t}}, 
\end{equation} 
or 3.46 utils per month with square-root utility. 

The change in the rate of change of the expectation value of the shipowner's utility 
resulting from entering into the insurance contract is
\begin{equation}
\gls{delta_ru_own}=\gls{ru_own_in}-\gls{ru_own_un},
\elabel{delta_ru_own}
 \end{equation} 
 or 0.094 utils per month with square-root utility. In the example with square-root utility the shipowner 
 should sign the contract, according to the utility paradigm.

Depending on the shipowner's utility function, $\Uo$, and his wealth, $\Wo$, buying insurance may increase or 
decrease the rate of change of the expectation value of his utility, with the result 
that the utility paradigm cannot say whether the contract should be signed or not. 
For almost 300 years this has been the celebrated classic resolution of the puzzle: 
since utility theory doesn't know what to do, it does not explicitly rule out the 
existence of insurance contracts.\\

\noindent\underline{The insurer's perspective}\\

It has been said that the utility paradigm applies to 
individuals only, and not to companies. For example, \citet[p.~631]{RothschildStiglitz1976} 
write: {\it ``It is less straight-forward to describe how insurance companies decide which 
contracts they offer for sale [...]. We assume that companies are risk neutral and that 
they are concerned only with expected profits.''} They continue in footnote 3: 
{\it ``the theory of the firm behaviour under uncertainty is one of the more 
unsettled areas of economic theory, we cannot look to it for the sort of support 
[...] which the large body of literature devoted to the expected utility theorem provides.''}

For reasons that will become clear in Sections~\ref{Solution_in_the_time} and  
\ref{The-infinite}, we don't follow \citet{RothschildStiglitz1976}. 
Instead, we assign a utility function to the insurer, a procedure followed 
by \citet{Bernoulli1738}. To do this we assign to the insurer a large wealth, 
$\gls{W_in}=\$1,000,000$. We simply treat the insurer as an individual, albeit with considerable means, and leave unspecified whether this is a natural person, a firm, or some other entity.

{\it Without} insurance, then, the (certain) rate of change of the insurer's utility is
\begin{equation} 
\gls{ru_in_un}=0
\end{equation} 
 (the insurer does no business).

{\it With} insurance, the rate of change of the expectation value of the insurer's utility is 
\begin{equation} 
\gls{ru_in_in}=\frac{ (1-\p)\Ui(\gls{W_in}+\F)+\p\Ui(\gls{W_in}+\F-\gls{L})-\Ui(\gls{W_in})}{\gls{Delta_t}},
\elabel{ru_in_in}
\end{equation} 
or 0.043 utils per month with square-root utility.

The change in the rate of change of the expectation value of the
insurer's utility resulting from entering into the contract is 
\begin{equation}
\gls{delta_ru_ins}=\gls{ru_in_in}-\gls{ru_in_un},
\elabel{delta_ru_ins}
\end{equation}
0.043 utils per month with square-root utility. The different perspectives are 
summarised in Table~\ref{utility_table}. Without specifying the utility functions 
the utility paradigm only makes weak 
predictions of behavioural consistency, based on assumed monotonicity 
(more money is better) and concavity (risk aversion) of the utility functions, 
as famously shown by \citet{vonNeumannMorgenstern1944}.

\begin{table}
\label{utility_table}
\noindent\makebox[\textwidth]{
\begin{tabular}{m{2cm} m{8cm} m{6.5cm}}
\gls{ru} & Shipowner & Insurer \\
\hline
insured & 
$\frac{\Uo(\Wo+\gls{G}-\F)-\Uo(\Wo)}{\gls{Delta_t}}$ &
$\frac{ (1-\p)\Ui(\gls{W_in}+\F)+\p\Ui(\gls{W_in}+\F-\gls{L})-\Ui(\gls{W_in})}{\gls{Delta_t}}$\\
uninsured &$\frac{(1-\p) \Uo(\Wo+\gls{G})+\p \Uo(\Wo-\C)- \U(\Wo)}{\gls{Delta_t}}$ &$0$\\
\hline
difference &$\frac{\Uo(\Wo+\gls{G}-\F)-(1-\p)\Uo(\Wo+\gls{G})-\p\Uo(\Wo-\C)}{\gls{Delta_t}}$ &$\frac{ (1-\p)\Ui(\gls{W_in}+\F)+\p\Ui(\gls{W_in}+\F-\gls{L})-\Ui(\gls{W_in})}{\gls{Delta_t}}$\vspace{.1cm}
\\
\hline
\hline
\end{tabular}
}
\caption{Changes in the expectation value of utility for shipowner and insurer}
\end{table}

 
\subsection{Solution in the time paradigm}
\label{Solution_in_the_time}

\subsubsection{Introducing time -- 20th- {\it vs.}~17th-century mathematics}

As \citet{Peters2011b} has argued,
changes in the expectation value of wealth are not {\it a priori} relevant to 
an individual. The expectation value is a mathematical object that has 
nice linearity properties which make it convenient for computations, 
but mathematical convenience is no reason for the object to be relevant. 
The physical operation that the expectation value encodes is
this: take all possible events (the ship is lost or not), and create
an ensemble of systems in numbers proportional to the
probabilities of the events that occur in them. For example, create
100 systems. Let the ship travel safely in 95 of them and let it be 
lost in the other five. Now take the different owners' wealths in all of these systems,
pool them, and share them equally among the owners. In this setup, each shipowner
receives with certainty $\G-\p\gls{L}$, \ie his gain minus the net premium.\footnote{We note similarities between this imagined setup and the practice -- now known as the ``general average'' -- developed around 800 BC by the inhabitants of Rhodes, where losses arising from the partial jettisoning of cargo at sea were "made good by the assessment of all [\dots] for the benefit of all" \citep[p.~164]{Scott1932}.}

Strictly speaking, this operation is an average over parallel
universes (the different systems). Less strictly speaking, the
operation corresponds to a large group of owners who make a
contract with each other, whereby, at the end of the month, all
wealths are pooled together, and shared 
equally.\footnote{See \citep{PetersAdamou2015a} for a discussion 
of the same operation, carried out repeatedly, in the context 
of the evolution of cooperation.}

Thus, the operation of taking the expectation value is in
essence the operation of signing an insurance contract -- a perfect
contract, no less, where all risk is eliminated and the premium is the
net premium. No wonder, then, that in the expected-wealth paradigm 
there is no good reason for signing an insurance contract -- 
mathematically, this has already happened!

Whichever way one chooses to conceptualise the expectation value, 
the central message should be clear: the rate of change of an 
individual's expected wealth bears no resemblance to, and is 
therefore a poor model of, how his wealth will evolve over time. 
The two mental pictures -- many parallel cooperating trajectories 
\textit{versus} a single trajectory unfolding over a long period -- 
are at odds.
In general we cannot
equate the performance of expectation values 
with the performance of a single system over time.

With this knowledge we can revisit the problem: clearly, an individual
shipowner will not come to sensible conclusions if he compares his
expectation value without buying insurance (which looks like he has signed 
a perfect insurance contract at the minimum fee that could possibly be offered) 
to his expectation value with buying insurance (which must be worse unless the insurer has 
made a mistake). As we will see, he {\it will} come to sensible conclusions 
if he uses the time-average growth rate of his wealth as a decision criterion.


\subsubsection{Solution}
Having argued that the treatment that led to the insurance puzzle is
invalid because it assumes that humans optimise over imagined copies of
themselves in parallel universes, it remains to show that the alternative
treatment resolves the puzzle. To be specific: the alternative
treatment is a different model of human decision making. The
traditional models say that humans make decisions by optimising the
expectation values of their wealth or utility. Our model says that humans make
decisions by optimising the time-average growth rates of their
wealth. For concreteness we assume multiplicative
dynamics.\footnote{Multiplicative dynamics is a simple and robust null
  model that solves the problem. We note that our treatment is
  not restricted to this special case but for clarity of exposition we refrain 
  from a more comprehensive treatment.}
Which model of human behaviour is better is only partly an empirical 
question.\footnote{We explicitly disagree with \citep{Friedman1953} on this point: 
the faithful reproduction of observed behaviour, even the 
successful prediction of not-yet observed behaviour, does not absolve a model from
the requirement of using reasonable assumptions. Prediction and understanding are 
not the same thing. \citet[p.~316]{RosenbluethWiener1945} point out
that {\it ``Not all scientific questions are directly amenable to experiment.''}} Our model seems 
{\it a priori} more plausible to us because it does not
introduce parallel universes, and because it resembles mechanistically the real-world scenario it seeks to reflect. 
It is empirically strong because it resolves the insurance puzzle, and conceptually 
strong because all elements of the solution have clear physical significance.

Both the utility paradigm and the time paradigm
consider the problem as it is stated to be underspecified. The utility paradigm assumes that 
people mentally convert money non-linearly into usefulness and then average arithmetically 
over all possible imagined outcomes; the time paradigm assumes that people mentally 
consider the future beyond the end of the single cargo shipment and are interested in the growth of their wealth. Neither assumption
is stated in the problem and each leads to a different type of solution, in which different 
choices must be made about the necessary additional information.

This additional information corresponds to introducing an axiom. The time paradigm
axiomatically assumes a stochastic dynamic of wealth. The utility paradigm axiomatically
assumes a utility function. The utility function, which encodes human behaviour is further
down the deductive chain -- it can be derived from the dynamic. Therefore the time paradigm
operates at a deeper level, and solutions in the time paradigm must be 
considered more fundamental. 
Another advantage of the time paradigm is that its axioms are more readily 
amenable to empirical verification: it is relatively easy to find out to what extent wealth is 
described by some dynamical model, whereas it is difficult to assess the 
value humans ascribe psychologically to changes in their wealth.

The time paradigm postulates that humans optimise the time-average growth 
rate of their wealth. The usual procedure for computing such a time average 
is to first transform wealth in such a way as to generate an ergodic observable 
and then compute the expectation value of that observable. Being the 
expectation value of an ergodic observable, this will be
the time average growth rate \citep{PetersGell-Mann2016}. Assuming
multiplicative repetition, the time average growth rate is
\begin{align}
\gls{g_time}&=\lim_{k\to\infty} \frac{1}{k \gls{Delta_t}}
\ln\left(\frac{\W(\gls{t}+k \gls{Delta_t})}{\W(\gls{t})}\right)\nonumber\\
&=\frac{\ave{\Delta \ln \W}}{\gls{Delta_t}}\nonumber\\
&=\frac{\ave{\ln\W(\gls{t}+\gls{Delta_t})}-\ln\W(\gls{t})}{\gls{Delta_t}},
\elabel{time_paradigm}
\end{align}
where the equality holds with probability 1. \Eref{time_paradigm} is
simply the expectation value of the exponential growth rate, which is
an ergodic observable that converges to a simple number under
multiplicative dynamics, see \citep[Chapter 7.7]{Gray2009} for further
discussion. Intriguingly, \eref{time_paradigm} is identical to 
\eref{ru} if we use logarithmic utility. Note that it is unnecessary to assume 
different utility functions for shipowner and insurer. The same logarithmic utility 
function for both makes the contract appear attractive to both according 
to the utility paradigm. This corresponds to both parties being subjected to 
the same dynamics. It turns out that utility functions 
correspond to specific dynamics, and the logarithm is twinned with 
multiplicative dynamics -- the most ubiquitous dynamic in growing and 
evolving systems. The two paradigms are fundamentally different, however. 
For instance, in the utility paradigm if observed behaviour is found to be 
incompatible with logarithmic utility, the investigation ends. One simply 
concludes that the actors involved have a different utility function, but this 
does not point the way to a deeper understanding of the processes 
involved. On the other hand, in the time paradigm, if observed behaviour 
is found to be incompatible with multiplicative dynamics, this prompts the 
question whether the prevailing dynamic may be a different one and 
why that may be, or whether the actor may be making 
bad decisions.

Once more we go through the two parties' perspectives.\\

\noindent\underline{The shipowner's perspective}\\

{\it Without} insurance, the time-average growth rate, under
multiplicative dynamics, of the shipowner's wealth is
\begin{align}
\gls{g_own_un}=\frac{1}{\gls{Delta_t}} \left((1-\gls{p})\ln\left(\frac{\gls{W_own}+\gls{G}}{\gls{W_own}}\right) +\p \ln\left(\frac{\Wo-\C}{\Wo}\right)\right),
\end{align}
or 1.9\% per month in our example. 

{\it With} insurance, the time-average growth rate is
\begin{equation}
\gls{g_own_in}=\frac{1}{\gls{Delta_t}} \ln\left(\frac{\gls{W_own}+\G-\gls{F}}{\gls{W_own}}\right),
\end{equation}
namely 2.2\% per month in our example. 

The change in the shipowner's time-average growth
rate of wealth, resulting from entering into the insurance contract,
is 
\begin{equation}
\gls{delta_g_own}=\gls{g_own_in}-\gls{g_own_un},
\end{equation}
here $+0.24\%$ per month.

Hence, the shipowner  should sign the contract, according to 
the time paradigm.
The rate of change of the expectation value of the shipowner's
wealth decreases, \eref{delta_r_own}, but that's irrelevant
because the shipowner is not an ensemble of owners sharing resources.\\

\noindent\underline{The insurer's perspective}\\

The only difference between the insurer and the shipowner (apart from
being at opposite sides of the contract) is in their wealths. Let's assume, 
as in Section~\ref{Solution_in_the_utility}, that the insurer's wealth 
is $\gls{W_in} = \$1,000,000$, namely ten times that of the shipowner.

{\it Without} insurance, the time-average growth rate of the insurer's
wealth is
\begin{align}
\gls{g_in_un}=0
\end{align}
because the insurer has no business.

{\it With} insurance, the time-average growth rate of the insurer's
wealth is
\begin{equation}
\gls{g_in_in}=\frac{1}{\gls{Delta_t}} \left( (1-\gls{p}) 
\ln\left(\frac{\gls{W_in}+\gls{F}}{\gls{W_in}}\right) +
\gls{p} \ln\left(\frac{\gls{W_in}+\gls{F}-\gls{L}}{\gls{W_in}}\right)\right),
\end{equation}
or $0.0071\%$ per month in our example. The difference in time-average
growth rates for the insurer is 
\begin{equation}
\gls{delta_g_ins}=\gls{g_in_in}-\gls{g_in_un},
\elabel{delta_g_ins}
\end{equation}
here $+0.0071\%$ per month. Hence, in the time paradigm not only the 
shipowner but also the insurer should sign the contract. We consider this

\mbox{}
\begin{mdframed}[linewidth=2pt]
\underline{The fundamental solution to the insurance puzzle:}\\
\\
\textbf{Both the shipowner and the insurer should sign the insurance contract, 
because this increases the time-average growth rates of both of their wealths.}
\end{mdframed}
\mbox{}

This is consistent both with the observed existence of insurance contracts 
and with human instincts about risk mitigation. It echoes Arrow's general observation that {\it ``there is always the simple justification for any contract freely arrived at between two individuals: if both of them choose to enter the contract, then both of them must be better off''} \citep[p.~137]{Arrow1971}.

Because we've assumed the mental model of the dynamics to be purely
multiplicative, only the relative risks matter: $\gls{L}$, $\gls{F}$, $\G$ and $\C$ only
appear as proportions of current wealth $\gls{W_own}$ or $\gls{W_in}$. 
In this sense the value of a dollar 
is indeed inversely proportional to the wealth of the decision maker, as 
\citep{Bernoulli1738} suggested on intuitive grounds. However, in 
the time paradigm the value of a dollar is given by the dynamic (the 
mode of repetition of the venture), not by psychology.

The insurance contract is seen to be beneficial to the shipowner 
and to the insurer -- a mathematical impossibility under the
expected-wealth paradigm. By taking time averages we 
find that the contract is a win-win situation, not a zero-sum game. 
This is possible because both parties are part of a non-stationary 
growth process. 
The utility paradigm often treats firms and individuals 
separately, see our quotes of \citep{RothschildStiglitz1976} in 
Section~\ref{Solution_in_the_utility}, but this is not necessary. 
We summarise the two perspectives in Table~\ref{time_table}, and illustrate the 
effect in \fref{time_fee}.

\begin{table}
\label{time_table}
\noindent\makebox[\textwidth]{
\begin{tabular}{m{2cm} m{7cm} m{6.5cm}}
$\gls{g_time}$ & Shipowner & Insurer \\
\hline
insured & 
$\frac{1}{\gls{Delta_t}} \ln\left(\frac{\gls{W_own}+\G-\gls{F}}{\gls{W_own}}\right)$&
$\frac{1}{\gls{Delta_t}} \left[ (1-\gls{p})
\ln\left(\frac{\gls{W_in}+\gls{F}}{\gls{W_in}}\right) +
\gls{p} \ln\left(\frac{\gls{W_in}+\gls{F}-\gls{L}}{\gls{W_in}}\right)\right]$\\
 uninsured &
$\frac{1}{\gls{Delta_t}} \left[(1-\gls{p})\ln\left(\frac{\gls{W_own}+\gls{G}}{\gls{W_own}}\right) +\p \ln\left(\frac{\Wo-\C}{\Wo}\right)\right]$&
0\\ \hline
difference &
$\frac{1}{\gls{Delta_t}} \left[ \ln\left(\frac{\gls{W_own}+\G-\gls{F}}
{(\gls{W_own}+\gls{G})^{(1-\gls{p})}(\gls{W_own}-\C)^{\gls{p}} }\right)\right]$
&
$\frac{1}{\gls{Delta_t}} \ln \left(\frac{(\gls{W_in}+\gls{F})^{(1-\gls{p})} (\gls{W_in}+\F-\gls{L})^{\gls{p}}}{\gls{W_in}}\right)$\\
\hline \hline
\end{tabular}
}
\caption{Time-average growth rates for shipowner and insurer}
\end{table}

Unlike in Table~\ref{expectation_table}, the bottom row of
Table~\ref{time_table} is not anti-symmetric -- the change in
time-average growth rate that the shipowner experiences when the
contract is signed is {\it not} the negative of the insurer's change. A price 
range exists where both parties gain from entering into the contract. Here 
systemic risk is reduced and systemic growth supported: both parties will
do better in the long run, which constitutes an explanation of the existence of an
insurance market.

\begin{figure}
\setlength{\unitlength}{\textwidth}
\begin{picture}(1,0.72)
\put(0,0){\includegraphics[width=1.\textwidth]{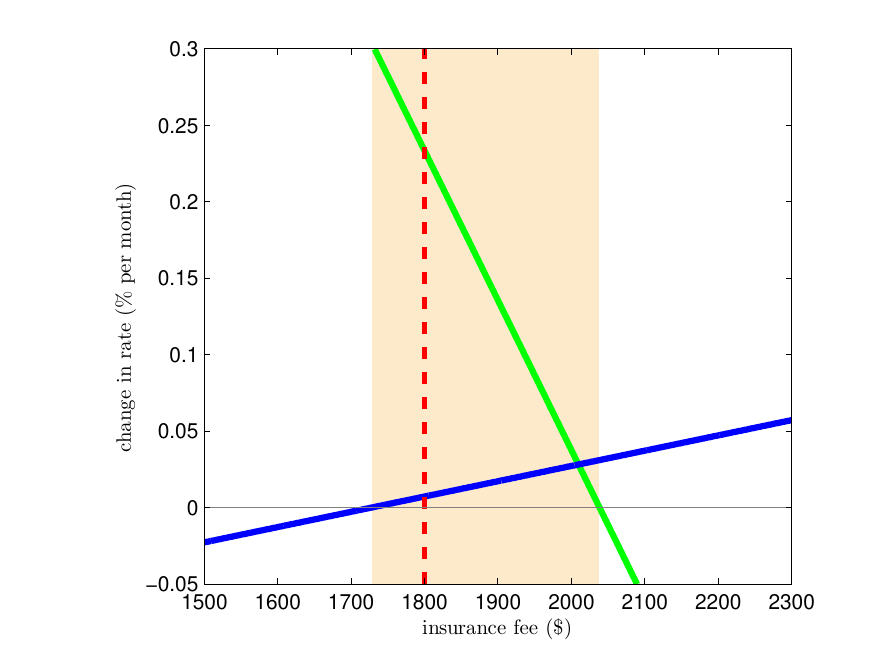}}
  \put(0.55,0.42){\includegraphics[width=0.35\textwidth]{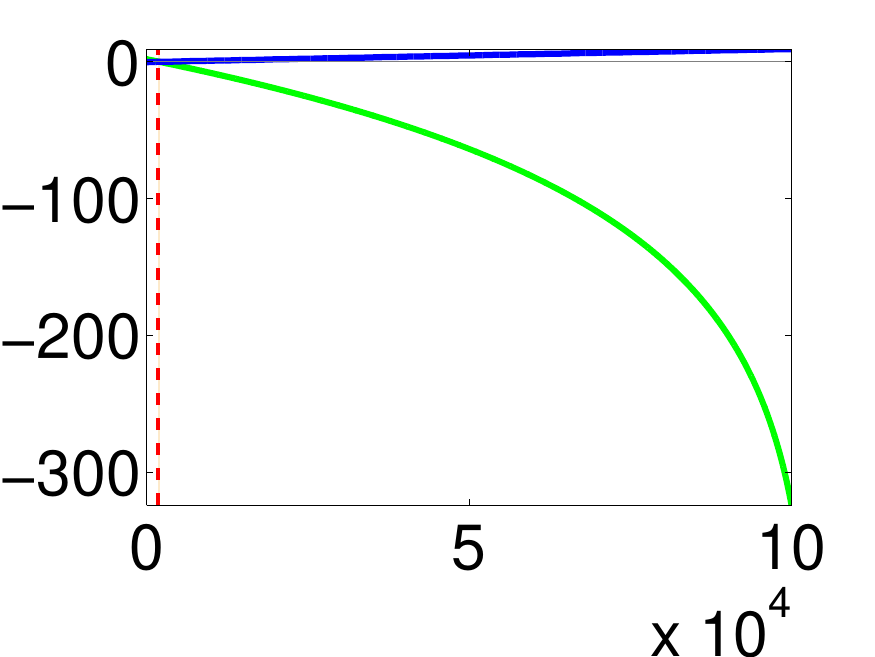}}
\end{picture}
\caption{Change in time-average growth rate resulting from signing an
  insurance contract versus insurance fee. \underline{Solid line:} 
  shipowner. \underline{Dashed line:} insurer. The vertical dotted line 
  indicates the proposed insurance fee. 
  \underline{Shaded region:} Averaging over appropriate ergodic observables 
  and without appealing to utility functions, a price range is seen to exist that is
  beneficial to both parties. Both the solid and dashed 
  lines lie above zero. What is typically defined as
  the ``fair price,'' the net premium, is detrimental to any insurer of 
  finite wealth (the dashed line is below zero at $\$1,700$ per month), leading to bankruptcy with
  probability 1. \underline{Inset:} horizontal axis extended to \$100,000. The non-linearity
  is clearly visible at these scales.} 
  \flabel{time_fee}
\end{figure}


\section{Discussion -- convergence to the expectation value in the large wealth limit}
\label{The-infinite}
The time solution to the insurance puzzle can also be presented by
considering the typical situation of an actual insurance company, such
as a car insurer. Such a company may insure millions of cars, and
therefore have access to a real-world ensemble that has nothing to do
with parallel universes, just different cars.

If the pool of insured cars is large enough, the rate at which the
insurer loses money to claims will be well described by the net
premium, $(\gls{F}-\gls{p}\gls{L})/\gls{Delta_t}$. This thought connects the time
paradigm to the expected-wealth paradigm. Imagine an insurer whose
wealth is much larger than any individual insured loss and any individual 
premium. In this case $\gls{delta_g_ins}$ of \eref{delta_g_ins} due to a single 
contract can be approximated by the first term of a Taylor expansion
\begin{align}
\gls{delta_g_ins}&= \frac{1}{\gls{Delta_t}} \left[(1-\gls{p})\ln \left(1+\frac{\gls{F}}{\gls{W_in}}\right)+\gls{p}\ln\left(1+\frac{\gls{F}-\gls{L}}{\gls{W_in}}\right)\right]\nonumber\\
&= \frac{1}{\gls{Delta_t}} \left[\frac{\gls{F}-\gls{p}\gls{L}}{\gls{W_in}} + o\left(\frac{\gls{F}}{\gls{W_in}}, \frac{\gls{F}-\gls{L}}{\gls{W_in}}\right)\right],
\elabel{convergence}
\end{align}
where we have used little-o notation. 
Multiplying the first-order term of  \eref{convergence} 
with $\gls{W_in}$ (an irrelevant constant at this order) yields exactly 
$\gls{delta_r_ins}$ of \eref{delta_r_ins}. In other words, as the risk that is being 
insured becomes negligible to the insurer, the predictions of the time 
paradigm for the insurer's behaviour become increasingly similar to those of the expectation-value 
paradigm. The two mathematical models of human behaviour are
identical in this limit, just as the models of Newtonian and
Einsteinian mechanics are identical in the limit of small velocities,
or the models of classical and quantum mechanics are identical in the
limit as Planck's constant approaches zero (the
``Correspondence Principle''). Similar behaviour was noted in the
context of leveraged investments in continuous time: in the small-leverage 
limit, the rate of fractional changes of the expectation value converges to the 
time-average exponential growth rate \citep{Peters2011a}.

This explains why the expectation-value criterion is sometimes a reasonable
description of large insurers' behaviour. Indeed, the assumption of risk neutrality 
for the insurance firm in \citep{RothschildStiglitz1976} is consistent with 
the time paradigm under multiplicative dynamics in the limit where the 
insurance fee and the insured loss are small compared to the wealth 
of the insurer. Outside this regime insurers that {\it ``are concerned only 
with expected profits''} \citep[p.~631]{RothschildStiglitz1976} 
will sooner or later go bankrupt because of excessive risk-taking. Any contract with $\F>\gls{p}\gls{L}$ 
and $\gls{L}>\Wi+\F$ will appear attractive to a Rothschild-Stiglitz insurer ($\gls{delta_r_ins}>0$) 
but the loss will bankrupt the firm when it occurs ($\gls{delta_g_ins}\to -\infty$).

Consequently, the behaviour outside 
the regime where the total underwritten risk is negligible compared to the insurer's resources 
is not well described by the maximisation of the expectation value of profits. 
Instead an actual insurance firm (or other risk-bearer) will set a limit for the acceptable 
probability of its bankruptcy over a given time interval, \ie a risk tolerance, and explore 
the possible behavioural choices within the resulting constraints, see 
\eg \citep{KaasETAL2008}. 

The difference apparent to \citep{RothschildStiglitz1976} between the shipowner (appearing to act 
according to a non-linear, concave, utility function) and the insurer (appearing risk neutral, \ie appearing 
to act according to a linear utility function) is simply a consequence of scale. At large wealth the possible 
logarithmic changes in wealth are small enough for a linear approximation to the logarithm to be valid. 
The insurer of infinite wealth can accept the net premium (the
ensemble average of the insured loss per time unit). The ensemble average 
is thus a limit that is often of practical relevance.

\section{Conclusion -- business is when both parties gain}

\Fref{time_fee} illustrates a remarkable difference to traditional
actuarial thinking in economics. According to the time paradigm there is a 
win-win regime where both parties gain (both the solid and dashed lines are above zero). The net premium, often called the ``fair price'' plays a 
less significant role -- a range of prices is beneficial for both parties. The 
fair price is never part of that range (it is the end of the range in the limit 
of the infinitely wealthy insurer). Where exactly the price of trade will lie 
can be negotiated. 

This seems to be a more general result: if it is true that people choose 
based on the physically sensible criterion of optimising growth over 
time, then a win-win range of prices will exist for any service or product 
that is traded. Business deals happen because both parties gain. This is the opposite 
of equilibrium thinking in economics, which tells us that deals
happen either because one party cons or coerces the other into an agreement, or
because both parties found the exact price where neither has a reason
not to get involved.

\section*{Acknowledgments}
We acknowledge a debt of gratitude to K.~Arrow for the discussions at the 30$^{\text{th}}$-anniversary symposium of the Santa Fe Institute which prompted this study. This manuscript is dedicated to his memory. We thank G.~Fulcher for helpful comments on an earlier version.

\section*{Funding}
This research did not receive any specific grant from funding agencies in the public, commercial, or not-for-profit sectors.

\section*{References}
\bibliographystyle{apalike}
\bibliography{bibliography}

\printnoidxglossaries

\end{document}